\documentclass[aps,prl,floatfix,twocolumn,showpacs,preprintnumbers,amsmath,amssymb,a4paper,superscriptaddress]{revtex4-1}

\usepackage{graphicx}
\usepackage{xcolor}
\usepackage{wasysym}
\usepackage{bm}

\begin{document}

\title{Dipolar confinement-induced resonances of ultracold gases in waveguides}

\author{P. Giannakeas}
\email{pgiannak@physnet.uni-hamburg.de}
\affiliation{Zentrum f\"{u}r Optische Quantentechnologien, Universit\"{a}t Hamburg, Luruper Chaussee 149, 22761 Hamburg,
Germany,}

\author{V.S. Melezhik}
\email{melezhik@theor.jinr.ru}
\affiliation{Bogoliubov Laboratory of Theoretical Physics, Joint Institute for Nuclear Research,
Dubna, Moscow Region 141980, Russian Federation,}

\author{P. Schmelcher}
\email{pschmelc@physnet.uni-hamburg.de}
\affiliation{Zentrum f\"{u}r Optische Quantentechnologien, Universit\"{a}t Hamburg, Luruper Chaussee 149, 22761 Hamburg,
Germany,}
\affiliation{The Hamburg Center for Ultrafast Imaging, Luruper Chaussee 149, 22761, Hamburg, Germany,}

\date{\today}

\begin{abstract}
We develop a non-perturbative theoretical framework to treat collisions with generic anisotropic interactions in quasi-one-dimensional geometries.
 Our method avoids the limitations of pseudopotential theory allowing to include accurately long-range anisotropic interactions. 
Analyzing ultracold dipolar collisions in a harmonic waveguide we predict dipolar confinement-induced resonances (DCIRs) which are attributed to different angular momentum states.
The analytically derived resonance condition reveals in detail the interplay of the confinement with the anisotropic nature of the dipole-dipole interactions.
The results are in excellent agreement with ab initio numerical calculations confirming the robustness of the presented approach.
The exact knowledge of the positions of DCIRs may pave the way for the experimental realization e.g. Tonks-Girardeau-like or super-Tonks-Girardeau-like phases in effective one-dimensional dipolar gases.
\end{abstract}


\maketitle
 In low-dimensional geometries due to tightly confining traps, ultracold atomic scattering undergoes crucial modifications yielding the effect of confinement-induced resonances (CIRs) \cite{olshanii98,rev}. 
A CIR is a Fano-Feshbach-type of resonance occurring when the scattering length $a_s$ and the length of the transversal confinement $a_\perp$ are comparable, namely $a_s/a_\perp\approx0.68$.
Remarkably, the deepened theoretical understanding of CIR physics \cite{nishida10, shlyapnikov01, blume04, kim06, melezhik09, sala11, saeidian08, gian12} has lead to major achievements in the experimental manipulation \cite{expecir0,expecir1} of interacting gaseous atomic matter.
Together with the extensive study of free-space dipolar collisions \cite{marinesku98, melezhik01, ticknor08, greene12, bohn09,deb01, roudnev09,yi01}, confinement-induced resonant scattering introduces an intriguing perspective for the control of dipolar many-body phases \cite{baranov08,car13}.
Indeed, reduced dimensionality has lead to the prediction of dipolar crystals \cite{buchler07} and the control of internal and external degrees of freedom of molecules has allowed the realization of dense ultracold polar molecule gases \cite{expe}.
In view of the substantial theoretical effort made on confined dipolar scattering \cite{kanjilal08,blume07,santos07,ticknor10,incao11,hanna12}, the need for a rigorous understanding of the role of anisotropic forces in CIRs becomes evident.

In this letter, we analytically derive the resonance condition for $s$-wave dipolar CIR (DCIR), with explicit dependence on the dipole-dipole interaction (DDI) strength.
This is done within an extended $K$-matrix formalism for harmonic quasi-one-dimensional (Q1D) geometries \cite{gian12} which incorporates anisotropic forces, i.e. the DDI, and takes into account contributions from higher angular momentum states.
These $\ell$-wave states are firstly coupled due to the anisotropic nature of the DDI and secondly by the harmonic confinement, which leads to a rich resonance structure of the DCIRs.
The $\ell$-wave DCIRs appear in the vicinity of shape resonances which are properly taken into account within the $K$-matrix approach going beyond the effective one-dimensional pseudopotential theory \cite{santos07}.
Interestingly, this interplay between the confinement and the DDI leads to an intricate dependence of the $s$-wave DCIRs positions on the dipolar interaction strength.
The derived resonance condition thus reveals in detail the impact of the DDI anisotropy on the CIR effect and provides the necessary tool for the experimental control of the dipolar collisions in Q1D traps. 
The exact knowledge of the positions of DCIRs can be utilized for the realization of a dipolar version of the (super-) Tonks-Girardeau gas \cite{expecir0} providing different dynamics in the collective oscillations of the many-body phase \cite{pedri}.
Notably, the present theoretical treatment equally can be applied to other collisional systems either of bosonic or fermionic symmetry where anisotropic forces dominate.
This includes metastable alkaline-earth-metal atoms in magnetic fields which interact with quadrupole-quadrupole interactions \cite{derevianko03} or rare-earth atoms, e.g. Dy, Er \cite{lu11, aikawa12}.

In the following, we consider a system of two bosonic, nonreactive polar molecules which collide in a Q1D waveguide.
They are treated as perfect dipoles fully polarized by an external electric field along the waveguide axis $\hat z$.
The transversal confinement is induced by a two-dimensional (2D) harmonic potential yielding a separation of the center of mass and relative degrees of freedom.
The physics of the collisional processes is then completely described by the relative Hamiltonian $H=-\frac{\hbar^2}{2\mu}\nabla^2+\frac{\mu}{2}\omega_{\perp}^2 \rho^2+V_{\rm int}(\mathbf{r})$ expressed in cylindrical coordinates ${\bf r}=(\rho,\phi,z)$,
where $\mu$ denotes the reduced mass and $\omega_{\perp}$ is the confinement frequency.
The interaction potential is $V_{\rm int}(\mathbf{r})=V_{\rm sh}(\mathbf{r})+\frac{d^2}{r^{3}}[1-3(\hat{z}\cdot \hat{r})^2]$, where the short-range term $V_{\rm sh}$ is modeled by a Lennard-Jones (LJ) 12-6 potential, $V_{\rm sh}(\mathbf{r})=\frac{C_{12}}{r^{12}}-\frac{C_{6}}{r^{6}}$, and the second term describes the DDI where $d$ is the induced dipole moment.
The justification of the LJ-potential for the short-range behavior of the two nonreactive molecules lies in the separation of the energy scales for elastic ultracold collisions and inelastic {\it chemical} processes \cite{hanna12}.
The ranges of the LJ and DDI potentials are $l_{\text{vdW}}=(2\mu C_6/\hbar^2)^{\frac{1}{4}}$ and $l_d=\mu d^2/\hbar^2$, respectively.
As in Refs.\cite{blume04,gian12}, we consider the confining oscillator length $a_{\perp} \equiv \sqrt{h/\mu \omega_{\perp}}$ to be the larger length scale, i.e. $a_{\perp}\gg l_{\text{vdW}}, l_d$.

This condition separates the configuration space into three domains with respect to the relative distance $r$:

(I) $l_d,~l_{\text{vdW}} < r \ll a_{\perp}$: At small separation distances the interactions $V_{\rm int}({\rm r})$ dominate, so that the dipoles effectively experience a free-space collision of total energy $E = \hbar^2k^2/2 \mu$, with the according symmetry imposed.
In this region, the corresponding wave function can thus be efficiently expanded in the $\ell$-wave angular momentum eigenstates.
Note that, due to the azimuthial rotational symmetry of $H$, the quantum number $m$ is conserved, and will here be fixed to $m = 0$.
On the contrary, the angular momentum $\ell$ is not conserved due to the anisotropy of the DDI, which couples states with $\Delta\ell = \ell-\ell'= 2$.
The scattering information of $V_{\rm int}$ is then imprinted in a free-space $K$-matrix of tridiagonal and symmetric form in $\ell$-representation, with entries for even $\ell$ due to bosonic symmetry;
considering up to $g$-wave contributions, namely $\ell=4$, it reads
\begin{eqnarray}
 \underline{K}^{\rm 3D}=
\begin{pmatrix}
	  K_{ss} & K_{sd} & 0 \\
	  K_{ds} & K_{dd} & K_{dg} \\
	  0 & K_{gd} & K_{gg} \\
      \end{pmatrix}.
\label{1}
\end{eqnarray}
Since the waveguide confinement is not experienced in this region, the entries of $\underline{K}^{\rm 3D}$ are a measure of the distortion of the wave function only by the presence of $V_{\rm int}$.

(II) $a_{\perp} \ll r \rightarrow \infty$:
At large separations the cylindrical confinement prevails, such that the wave function decomposes into the radial eigenmodes $n$ of the transversal 2D harmonic oscillator.
These are regarded as asymptotic channels of the scattering process, which is described by the corresponding $K$-matrix in one dimension, $\underline{K}^{\rm 1D}$.

(III) $l_d,~l_{\text{vdW}} \ll r \ll a_{\perp}$: At intermediate separation distances there is, in general, an admixture of $\ell$-wave states from $V_{\rm int}$ and $n$-mode states from the confinement.
However, in this particular domain, both the interaction and confining potentials essentially vanish, so that the corresponding scattering solutions can be efficiently matched.
Therefore, a non-orthogonal {\it local frame transformation} $\underline{U}$ exists \cite{greene87}, which enables the projection of the wave function of domain (I) onto that of domain (II).
In particular, $\underline{U}$ transforms the corresponding $K$-matrices into one another: $\underline{K}^{1D}=\underline{U}^T ~\underline{K}^{3D}~\underline{U}$ with its elements being $(\underline{K}^{1D})_{nn'}=\sum_{\ell \ell'}U^T_{n\ell}K_{\ell \ell'}^{3D}U_{\ell'n'}$.
Note that both $\underline{K}^{3D}$ and $\underline{K}^{1D}$ depend on the total energy $E$.

In the following, we consider low-energy scattering in the asymptotic transversal ground state $n = 0$, following the criterion $q_0a_{\perp} \ll 1$ for the relative longitudinal wave vector $q_0$ which is considered to be small and finite.
The total collision energy, in domain II, is $E = \hbar^2k^2/2 \mu=\hbar\omega_{\perp}+\hbar^2q_0^2/2\mu$, so that only the first channel ($n=0$) is energetically open ($o$), while higher channels ($n>0$) remain closed ($c$), since $E<\hbar \omega_{\perp}(2n+1)$.
The asymptotic components of the wave function in the $c$-channels, however, contains exponential divergences, rendering the scattering {\it{unphysical}}.
This behavior is remedied in the framework of multichannel quantum defect theory (MQDT) by imposing the physically acceptable boundary conditions in $c$-channels \cite{greene96}, yielding a {\it{physical}} $K$-matrix given by
\begin{equation}
 \underline{\tilde{K}}_{oo}^{1D} = \underline{K}_{oo}^{1D}+i\underline{K}^{1D}_{oc}(\mathcal{I}-i\underline{K}^{1D}_{cc})^{-1}\underline{K}^{1D}_{co},
\label{dyson}
\end{equation}
where $\mathcal{I}$ is the identity matrix.
The roots of $\text{det}(\mathcal{I}-i\underline{K}^{1D}_{cc})=0$ provide the bound states in the $c$-channels which energetically lie in the continuum of the $o$-channels.

Using the Dyson-like form of $\underline{\tilde{K}}_{oo}^{1D}$ from Ref.~\cite{gian12}, we obtain
\begin{equation}
\underline{\tilde{K}}_{oo}^{1D}=-\frac{\boldsymbol{\alpha\cdot\mathcal{\xi}}}{{q_0 a_{\perp}(1+\boldsymbol{\alpha \cdot \mathcal{R}})}},
\label{kphys}
\end{equation}\\
where the scalar products are short-hand for summed coefficients of the $\boldsymbol{\alpha}, \boldsymbol{\xi}$ and $\boldsymbol{\mathcal{R}}$, given explicitly in the supplementary material (SM).
$\boldsymbol{\xi}$ and $\boldsymbol{\mathcal{R}}$ contain traces $\sum_n U^T_{n\ell}U_{\ell'n}$ over the $c$-channels $n$, yielding combinations of the Hurwitz $\zeta$-function for each ($\ell$, $\ell'$)-pair.
$\boldsymbol{\alpha}$ consists of single-term combinations of ${\bar a}_{\ell \ell^{'}} \equiv {a_{\ell \ell^{'}}}/{a_{\perp}}$, where $a_{\ell \ell^{'}}=-K_{\ell \ell^{'}}/k$ are generalized, energy dependent scattering lengths for dipolar collisions in free-space \cite{bohn09}.
Due to Eq.~(\ref{1}) we remark that $\boldsymbol{\alpha}$ contains up to $g$-wave $\bar{a}_{\ell\ell'}$-terms.
Eq.~(\ref{kphys}) therefore encapsulates directly the impact of the anisotropy of the DDI on the confined scattering:
$\underline{\tilde{K}}_{oo}^{1D}$ is determined by the dipole-induced $\bar{a}_{\ell \ell^{'}}$-terms contained in $\boldsymbol{\alpha}$, which are simultaneously weighted by the coupling due to the confinement via $\boldsymbol{\xi}$ and $\boldsymbol{\mathcal{R}}$.

The resonant behavior arises in the form of poles of $\underline{\tilde{K}}^{1D}_{oo}$, given by the roots of the equation
\begin{equation}
1+\boldsymbol{\alpha \cdot \mathcal{R}}=0.
\label{res_cond}
\end{equation}
These coincide with the zeros of $\text{det}(\mathcal{I}-i\underline{K}^{1D}_{cc})=0$ in Eq.~(\ref{dyson}), which demonstrates that the origin of the resonance structure is a Fano-Feshbach mechanism.
Note that the resonance condition, Eq.~(\ref{res_cond}), can be met in multiple ways by allowing either one of the $\bar {a}_{\ell \ell'}$-terms to be dominant.

It is thus evident that due to the presence of the DDI within the waveguide, different $\ell$-wave states from the domain (I) contribute to this resonance mechanism.
The $\ell$-wave labeling is still used in the sense that a particular partial wave dominates over the others, although they are coupled together due to DDI, which leads to broad $s$-wave and narrow $(\ell > 0)$-wave DCIRs with corresponding positions in the parameter space, e.g. $l_d$, $l_{\rm vdW}$, $a_{\perp}$, determined by Eq.~(\ref{res_cond}).
Note that this difference of the widths arises from the free-space dipolar scattering, where within the adiabatic approximation of the two-body dynamics \cite{roudnev09} the $s$-wave adiabatic channel possesses an effective  $-1/r^4$ potential tail, while each $\ell > 0$ state exhibits repulsive barriers leading to increasingly narrower resonances with increasing $\ell$-wave character.

We will now focus on the $s$-wave DCIRs, which are of immediate experimental relevance.
Thus, we isolate the free-space $a_{ss}$ dipolar scattering length on the left hand side of Eq.~(\ref{res_cond}), and we obtain the corresponding resonance condition ${\bar a}_{ss}(ka_{\perp},d) = \mathcal{F}(\{{\bar a}_{\ell\ell'}(ka_{\perp},d)\}, \{{\mathcal R}_i(ka_{\perp})\})$, where
\begin{widetext}
\begin{equation}
\mathcal{F} = - \frac{ 1 + \bar{a}_{dd} (\mathcal{R}_2+\bar{a}_{gg} \mathcal{R}_{12})
+\bar{a}_{gg} (\mathcal{R}_3+\bar{a}_{sd} \mathcal{R}_{11})
+\bar{a}_{sd} (\mathcal{R}_4+\bar{a}_{dg} \mathcal{R}_{13})
+\bar{a}_{dg} (\mathcal{R}_5+\bar{a}_{dg} \mathcal{R}_{7})
+\bar{a}_{sd}^2 (\mathcal{R}_6+\bar{a}_{gg} \mathcal{R}_{14})}{\mathcal{R}_1
+\bar{a}_{dd} \mathcal{R}_{10}
+\bar{a}_{gg} \mathcal{R}_{8}
+\bar{a}_{dg} \mathcal{R}_{9}
+(\bar{a}_{dg}^2-\bar{a}_{dd} \bar{a}_{gg} )\mathcal{R}_{14}},
\label{F}
\end{equation}
\end{widetext}
with the $\mathcal{R}_i$ ($i=1\ldots14$) being the components of $\boldsymbol{\mathcal{R}}$ (see SM). 
Originating from the poles of $\underline{\tilde{K}}_{oo}^{1D}$ in Eq.~(\ref{kphys}), the quantity $\mathcal{F}$ now embodies the interplay between the DDI anisotropy and the confinement.
Note that in general ${\bar a}_{ss}(ka_{\perp},d)$ and $\mathcal{F}(\{{\bar a}_{\ell\ell'}(ka_{\perp},d)\}, \{{\mathcal R}_i(ka_{\perp})\})$ depend differently on the dipole moment $d$, thusly their equality provides a transcendental equation being fulfilled for particular values of $d$ in the parameter space.

The $s$-wave DCIRs appear in the vicinity of the free-space resonances which are $s$-wave dominated, i.e. near the broad divergences of the $a_{ss}$ dipolar scattering length, where all the higher $a_{\ell\ell'}$ scattering lengths are non-resonant.
Additionally, as was shown in Ref.\cite{bohn09, blume07} within the Born approximation (BA) the non-resonant $a_{\ell\ell'}$, except $\ell=\ell'=0$ terms are proportional to the dipolar length $l_d$.
Consequently, this universal, i.e. $V_{sh}$-independent, behavior of the higher $a_{\ell \ell'}$ terms leads to the following simplification of Eq.~(\ref{F}):
\begin{equation}
\mathcal{F}_{\rm BA} = - \frac{ 1 + \eta_1\bar{l}_d + \eta_2 \bar{l}_d^2 + \eta_3\bar{l}_d^3 }{\sigma_0 + \sigma_1\bar{l}_d + \sigma_2 \bar{l}_d^2 },
\label{FBorn}
\end{equation}
where $\bar{l}_d \equiv l_d/a_{\perp}$ and the $\eta_j$ and $\sigma_j$ consist of combinations of ${\mathcal{R}}_i$'s (see SM).
In the considered low-energy limit $q_0a_{\perp}\ll1$, they acquire the values $\eta_1\approx1.844$, $\eta_2\approx-1.119$, $\eta_3\approx0.013$, $\sigma_0\approx-1.46$, $\sigma_1\approx2.008$ and $\sigma_2\approx0.046$.
For $l_d = 0$, the resonance condition, $\bar{a}_{ss} = \mathcal{F}_{\rm BA}$, reduces to $\bar{a}_{s} = - 1/\sigma_0 = 0.68$, as expected for the $s$-wave CIR \cite{olshanii98}.

\begin{figure}[b!]
\includegraphics[scale=0.70]{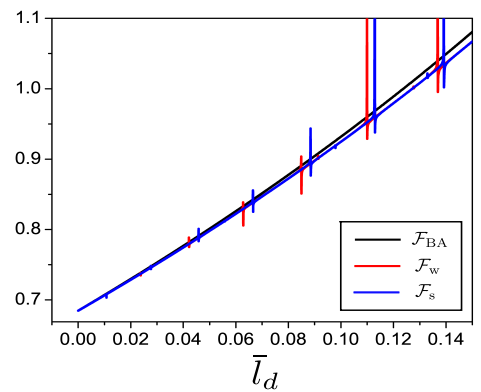}
\caption{ (color online) $\mathcal{F}$ versus $\bar{l}_d \equiv l_d/a_{\perp}$. The black line refers to $\mathcal{F}_{BA}$ (Eq.~(\ref{FBorn})), whereas $\mathcal{F}_{\rm w}$ (red line) and $\mathcal{F}_{\rm s}$ (blue line) refer to Eq.~(\ref{F}) where the terms $a_{\ell\ell'}$ are numerically calculated for a LJ potential that possesses for $d=0$ a weakly or a strongly $s$-wave bound state, respectively.}
\label{fig1}
\end{figure}

The analytical results from the BA, Eq.~(\ref{FBorn}), are compared in Fig.\ref{fig1} to those of Eq.~(\ref{F}), in which the $a_{\ell\ell'}$ are calculated numerically from the dipolar free-space problem \cite{melezhik91}.
We see that, whereas the BA breaks down close to free-space resonances ($a_{\ell\ell'}\to \infty$, $\ell,~\ell'\geq 0$) as expected \cite{taylor}, $\mathcal{F}_{\rm BA}$ is in good agreement with the non-resonant parts of $\mathcal{F}$;
these are also the regimes of interest here, since the $s$-wave DCIRs occur away from $a_{\ell\ell'}$, $\ell,~\ell'\geq 0$ free-space resonances.

\begin{figure}[t!]
\includegraphics[width=\columnwidth]{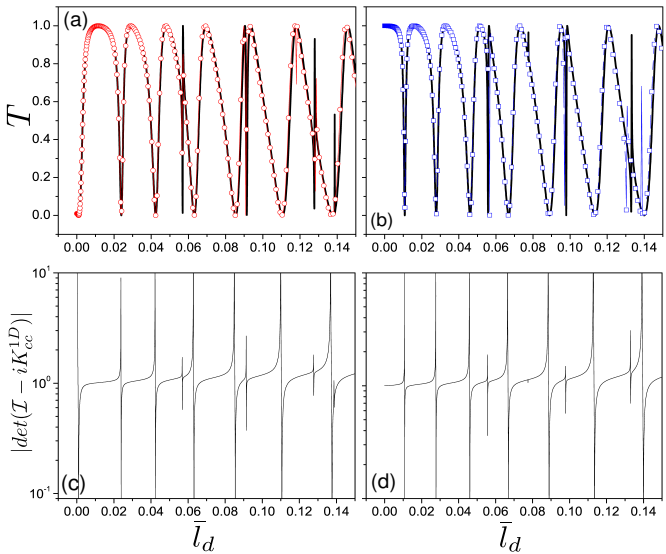}
\caption{ (color online) The transmission $T$: analytical results (solid line) and numerical calculations for (a) $a_s\gg l_{\rm{vdW}}$ ($\color{red}{-\ocircle -}$), (b) $a_s\ll l_{\rm{vdW}}$ ($\color{blue} {-\square-}$). (c) and (d) show the corresponding quantity $|det(\mathcal{I}-iK_{cc}^{1D})|$, for the parameter values of (a) and (b), respectively.}
\label{fig2}
\end{figure}

To study the universal aspects of $\mathcal{F}$, we consider for $d=0$ the two limiting cases of a weakly and a strongly $s$-wave bound state in the LJ potential, yielding $a_s \gg l_{\rm{vdW}}$ and $a_s\ll l_{\rm{vdW}}$, respectively.
We observe in Fig.~\ref{fig1} that $\mathcal{F}_{\rm s}$ (strongly bound-blue line) and $\mathcal{F}_{\rm w}$ (weakly bound-red line) practically coincide everywhere apart from the positions of the narrow resonant features.
The non-universality of these resonant features mainly arises from the coupling term $a_{sd}$, which is strongly affected by the non-universal and strongly resonant behavior of $a_{ss}$.
However, the distance between the corresponding divergences in $\mathcal{F}_{\rm s}$ and $\mathcal{F}_{\rm w}$ decreases as $\bar{l}_d$ increases.
This is because, as the DDI becomes stronger, it dominates the short-range LJ interaction and eventually shields it completely, thereby restoring the universal behavior of $a_{ss}$.

We now investigate the resonant structure of the transmission coefficient $T$ for confined dipolar scattering, which we derive analytically in terms of the physical $K$-matrix (see SM) as $T = [1+(\underline{\tilde{K}}^{1D}_{oo})^2]^{-1}$.
Note that the numerically calculated $\boldsymbol{\alpha}$ from the unconfined problem is used as an input in Eq.~(\ref{kphys}), as before.
The transmission is shown in Fig.\ref{fig2} as a function of $\bar{l}_d$, again for the cases of (a) a weakly and (b) a strongly bound $s$-wave state in the LJ potential at $d=0$, each featuring sequences of $s$- and higher partial waves ($\ell>0$) DCIRs.
As mentioned previously, the latter are much narrower due to the presence of the repulsive barrier.
These results are compared to exact numerical calculations of $T$ (red circles and blue squares in Fig.\ref{fig2} (a) and (b), respectively) based on a scheme presented in Ref.~\cite{melezhik12}.
An excellent agreement is observed.
Moreover, Fig.\ref{fig2} (c) and (d) show the corresponding expression $|det(\mathcal{I}-iK_{cc}^{1D})|$, which is observed to tend to zero exactly at the positions of $T \approx 0$.
This indeed illustrates that the $\ell$-wave DCIRs fulfill a Fano-Feshbach scenario.
\begin{figure}[t!]
\includegraphics[width=\columnwidth]{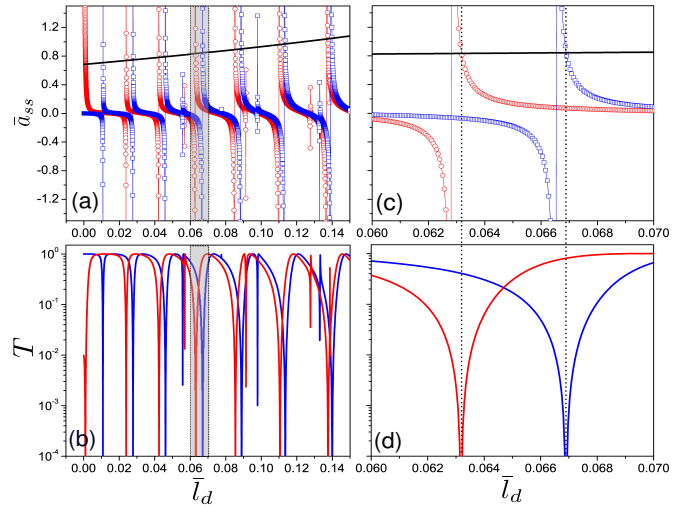}
\caption{ (color online) (a) $\bar{a}_{ss}$ depicted for $a_s\gg l_{\rm{vdW}}$ ($\color{red}{-\ocircle-}$) and $a_s\ll l_{\rm{vdW}}$ ($\color{blue} {-\Box-}$), whereas the black line shows the quantity $\mathcal{F}_{BA}$, (b) the analytically calculated transmission $T$ for $a_s\gg l_{\rm vdW}$ (red line) and $a_s\ll l_{\rm vdW}$ (blue line). (c) and (d) represent magnifications of the gray shaded areas of (a) and (b), respectively.}
\label{fig3}
\end{figure}

Fig.\ref{fig3} (a) shows a graphical solution of the resonance condition for the positions of the $s$-wave DCIRs: the dipolar scattering within the waveguide becomes resonant for the values of $\bar{l}_d$ where $\bar{a}_{ss} = \mathcal{F} \simeq \mathcal{F}_{\rm BA} > 0$.
We focus on the broad divergences of $\bar{a}_{ss}$ related to $s$-wave dominated dipolar free-space resonances, whose width increases with $\bar{l}_d$, since the DDI becomes more attractive.
The intersections of $\mathcal{F}_{\rm BA}$ and $\bar{a}_{ss}$ in Fig.\ref{fig3}~(a) are seen to occur exactly at the zeros $T(\bar{l}_d) = 0$ of the transmission in Fig.\ref{fig3}~(b), as clearly demonstrated for the magnified grey shaded regions in Fig.\ref{fig3} (c) and (d).
This striking coincidence demonstrates the high accuracy of the derived resonance condition for $s$-wave DCIRs, for both limiting cases $a_s \ll l_{\text{vdW}}$ and $a_s \gg l_{\text{vdW}}$.

This accurate prediction of the $s$-wave DCIRs is retained over the whole range of the DDI strengths, from the non-universal to the universal regime ($\bar{l}_d>0.11$) of the $s$-wave dominated dipolar free-space resonances.
In the universal regime the short-range physics represented by $V_{sh}$ indeed becomes irrelevant, in the sense that the positions of the $s$-wave DCIRs (intersections of $\bar{a}_{ss}$ and $\mathcal{F}_{\rm BA}(\bar{l}_d)$) for different strengths of $V_{\rm sh}$ to a good approximation coincide in the ($\bar{a}_{ss}, \bar{l}_d$)-plane.

Fig.\ref{fig3} (a) comprises the most intriguing difference of DCIRs compared to well-established CIRs with {\it isotropic} interactions:
In contrast to the case of CIRs, the positions of $s$-wave DCIRs, measured in values of $\bar{a}_{ss}$, increases for increasing interaction strength of the DDI potential, or, equivalently, for an increasing number of $s$-wave dominated dipolar free-space resonances which have become bound within $V_{\rm int}({\bf r})$.
Therefore, in the presence of DDI the confinement-induced shift of free-space resonances is no longer constant, but increases for successive resonances passing the open channel threshold.
This behavior arises from the anisotropic nature of the DDI, which mixes higher partial $\ell$-waves more strongly for larger $l_d$, and is enhanced via their recoupling by the confinement, yielding substantial contributions to the positions of the $s$-wave DCIRs.

In conclusion, we have extended the $K$-matrix formalism to treat dipolar collisions and include higher partial waves in the presence of a harmonic Q1D confinement leading to the prediction of $\ell$-wave DCIRs.
In particular, we analyzed in detail the case of $s$-wave DCIRs and derived analytically the corresponding resonance condition in the form $\bar{a}_{ss} = \mathcal{F}_{\rm BA}$, containing explicitly the dependence on the DDI strength.
Apart from providing an essential ingredient for the resonant control of nonreactive polar molecule gases in Q1D, this result sheds light on the physics underlying the DCIR effect:
We demonstrate how the DDI couples the $\ell$-wave states of the short-range potential, which are in turn mixed by the confinement.
We remark that in the case of fermionic dipolar collisions we expect the appearance of corresponding DCIRs, which may exhibit universal characteristics even for weak dipole moments as BA indicates \cite{bohn09}. 
In addition, for non-polarized dipoles, including the azimuthal $\phi$ dependence in our framework we may encounter $\ell$-wave DCIRs which split into components of the azimuthal quantum number $m$ \cite{ticknor}.
The theoretical advance presented here, combined with MQDT theory \cite{ruzik12}, may pave the way for new insights on reactive polar molecule collisions in quasi-2D \cite{expe}.
Furthermore, we remark that the s-wave dominated dipolar free-space resonances can be induced either by strong dc-electric or laser fields \cite{marinesku98,melezhik01, deb01}.
This together with the tunability of the confinement frequency provides us with excellent tools for probing s-wave DCIRs, which experimentally can be identified by their shifts for successive free-space dipolar resonances (see Fig.\ref{fig3}(a)).
Notably, as Fig.\ref{fig3}(a)-(b) indicate the density of the $s$-wave DCIRs does not or only weakly depend on the short-range physics, though it is strongly affected by the density of the free-space resonances (see \cite{hanna12}).
For a given density of free-space resonances the corresponding DCIRs might be very well experimentally resolvable, since their width increases as $\bar{l}_d$ increases (see Fig.\ref{fig2}(a)-(b)).

The authors thank I. Brouzos, C. Morfonios and F. K. Diakonos for valuable comments and discussions.
V.S.M. acknowledges financial support by the Deutsche Forschungsgemeinschaft and the Heisenberg-Landau Program.
P.S. acknowledges the Deutsche Forschungsgemeinschaft for financial support.

\end{document}


\title{Supplemental Material}

\author{P. Giannakeas}
\email{pgiannak@physnet.uni-hamburg.de}
\affiliation{Zentrum f\"{u}r Optische Quantentechnologien, Universit\"{a}t Hamburg, Luruper Chaussee 149, 22761 Hamburg,
Germany,}

\author{V.S. Melezhik}
\email{melezhik@theor.jinr.ru}
\affiliation{Bogoliubov Laboratory of Theoretical Physics, Joint Institute for Nuclear Research,
Dubna, Moscow Region 141980, Russian Federation,}

\author{P. Schmelcher}
\email{pschmelc@physnet.uni-hamburg.de}
\affiliation{Zentrum f\"{u}r Optische Quantentechnologien, Universit\"{a}t Hamburg, Luruper Chaussee 149, 22761 Hamburg,
Germany,}
\affiliation{The Hamburg Center for Ultrafast Imaging, Luruper Chaussee 149, 22761, Hamburg, Germany,}

\date{\today}

\maketitle
In the following we present the explicit form of the {\it {physical}} $K^{1D}$-matrix and the corresponding scalar products, as well as the relations of the $\eta$-constants which refer to the resonance condition and finally, we discuss the derivation of the transmission coefficient $T$.

\section{I. The explicit form of the physical $K^{1D}$-matrix for quasi-1D geometries}
Let us provide some insights into the derivation of the physical $\underline{K}^{1D}$-matrix, namely $\underline{\tilde{K}}_{oo}^{1D}$, for dipole-dipole collisions.
In general we have

\begin{equation}
 \underline{\tilde{K}}_{oo}^{1D} = \underline{K}_{oo}^{1D}+i\underline{K}^{1D}_{oc}(\mathcal{I}-i\underline{K}^{1D}_{cc})^{-1}\underline{K}^{1D}_{co},
\label{dyson}
\end{equation}
where we assumed that only one energetically open ($o$-) channel, namely $n=0$ and $n\neq0$ energetically closed ($c$-) channels.
Here $\underline{K}_{oo}^{1D}=\sum_{\ell \ell'}\underline{K}^{1D,\ell \ell'}_{oo}$ refers to the open-open channel transitions and the matrix element $\underline{K}^{1D,\ell \ell'}_{oo}$ is defined as $(\underline{K}^{1D,\ell\ell'}_{oo})_{00}= (U^T)_{0\ell}K^{3D}_{\ell \ell'}U_{\ell'0}$ with $K^{3D}_{\ell \ell'}$ being the corresponding matrix element of $\underline{K}^{3D}$.
$\underline{K}^{1D}_{oc}=\sum_{\ell\ell'} \underline{K}^{1D,\ell\ell'}_{oc}$ ($\underline{K}^{1D}_{co}=\sum_{\ell\ell'} \underline{K}^{1D,\ell\ell'}_{co}$) refers to open-closed (closed-open) transitions and each $\underline{K}^{1D,\ell\ell'}_{oc}$ ($\underline{K}^{1D,\ell\ell'}_{co}$) is a single row- (column-) matrix, where its entries are defined as $(\underline{K}^{1D,\ell\ell'}_{oc})_{0n}=(U^T)_{0\ell} K^{3D}_{\ell \ell'}U_{\ell'n}$ $\big($ $(\underline{K}^{1D,\ell\ell'}_{co})_{n0}=(U^T)_{n\ell}K^{3D}_{\ell \ell'}U_{\ell'0}$ $\big)$.
$\underline{K}^{1D}_{cc}=\sum_{\ell\ell'}\underline{K}^{1D,\ell\ell'}_{cc}$ refers to closed-closed transitions, and the matrix elements of $\underline{K}^{1D,\ell\ell'}_{cc}$ are given by $(\underline{K}^{1D,\ell\ell'}_{cc})_{nn'}=(U^T)_{n\ell}K^{3D}_{\ell \ell'}U_{\ell'n'} $where each $\underline{K}^{1D,\ell\ell'}_{cc}$ is a rank one matrix.

The matrix $(\mathcal{I}-i\underline{K}^{1D}_{cc})\equiv\mathcal{I}-i\sum_{\ell\ell'}\underline{K}^{1D,\ell\ell'}_{cc}$ is inverted analytically where for each $\underline{K}^{1D,\ell\ell'}_{cc}$ rank one matrix we use recursively the inversion identity for rank one matrices of Ref.~\cite{miller81}.
Thus $\underline{\tilde{K}}_{oo}^{1D}$ takes a Dyson-like form by rewriting  Eq.~(\ref{dyson}) in terms of $\underline{K}^{1D,\ell\ell'}_{oo}$, $\underline{K}^{1D,\ell\ell'}_{oc}$, $\underline{K}^{1D,\ell\ell'}_{co}$ and the analytically inverted  $\mathcal{I}-i\sum_{\ell\ell'}\underline{K}^{1D,\ell\ell'}_{cc}$  for each $\ell\ell'-$pair.
Thereafter, performing analytically all the corresponding matrix multiplications Eq.~(\ref{dyson}) then reads
\begin{equation}
\underline{\tilde{K}}_{oo}^{1D}=-\frac{\boldsymbol{\alpha\cdot\mathcal{\xi}}}{{q_0 a_{\perp}(1+\boldsymbol{\alpha \cdot \mathcal{R}})}},
\label{kphys}
\end{equation}
where the scalar product $\boldsymbol{\alpha \cdot \mathcal{R}}$ has the following form:

\begin{eqnarray}
\boldsymbol{\alpha \cdot \mathcal{R}}&=&\bar{a}_{ss}\mathcal{R}_{1}+\bar{a}_{dd}\mathcal{R}_{2}+\bar{a}_{gg}\mathcal{R}_{3}+
\bar{a}_{sd}\mathcal{R}_{4}+\bar{a}_{dg}\mathcal{R}_{5}+
\bar{a}_{sd}^2\mathcal{R}_{6}+\bar{a}_{dg}^2\mathcal{R}_{7}+\bar{a}_{ss}\bar{a}_{gg}\mathcal{R}_{8} \nonumber\\
&+&\bar{a}_{ss}\bar{a}_{dg}\mathcal{R}_{9}+\bar{a}_{ss}\bar{a}_{dd}\mathcal{R}_{10}+\bar{a}_{gg}\bar{a}_{sd}\mathcal{R}_{11}+
\bar{a}_{dd}\bar{a}_{gg}\mathcal{R}_{12}+
\bar{a}_{sd}\bar{a}_{dg}\mathcal{R}_{13} \nonumber\\
&+&(\bar{a}_{ss}\bar{a}_{dg}^2+\bar{a}_{gg}\bar{a}_{sd}^2-
\bar{a}_{ss}\bar{a}_{dd}\bar{a}_{gg})\mathcal{R}_{14},
 \label{1}
\end{eqnarray}
with ${\bar a}_{\ell \ell^{'}} \equiv {a_{\ell \ell^{'}}}/{a_{\perp}}$ with $a_{\ell \ell^{'}}=-K^{3D}_{\ell \ell^{'}}/k$ being generalized, energy dependent scattering lengths for dipolar collisions in free-space \cite{bohn09} encapsulating the scattering information of the short-range and the dipole-dipole interactions.
The coefficients $\mathcal{R}_i$, $i=1,\ldots,14$, are the components of the vector $\boldsymbol{\mathcal{R}}$ and are defined by the following:

\begin{eqnarray}
\mathcal{R}_{1}&=&b_1;~~\mathcal{R}_{2}=\frac{b_2}{(ka_{\perp})^4};~~\mathcal{R}_{3}=\frac{b_6}{(ka_{\perp})^8};~~
\mathcal{R}_{4}=\frac{2b_3}{(ka_{\perp})^2};~~\mathcal{R}_{5}=\frac{2b_4}{(ka_{\perp})^6};~~\mathcal{R}_{6}=\frac{b^2_3-b_1b_2}{(ka_{\perp})^4}; \nonumber\\
\mathcal{R}_{7}&=&\frac{b^2_4-b_6b_2}{(ka_{\perp})^{12}};~~\mathcal{R}_{8}=\frac{b_1b_6-b^2_5}{(ka_{\perp})^8};
~~\mathcal{R}_{9}=\frac{2(b_1b_4-b_3b_5)}{(ka_{\perp})^6};~~\mathcal{R}_{10}=-\frac{b^2_3-b_1b_2}{(ka_{\perp})^4}; \nonumber\\
\mathcal{R}_{11}&=&\frac{2(b_3b_6-b_4b_5)}{(ka_{\perp})^{10}};~~\mathcal{R}_{12}=-\frac{b^2_4-b_6b_2}{(ka_{\perp})^{12}};
~~\mathcal{R}_{13}=\frac{2(b_3b_4-b_2b_5)}{(ka_{\perp})^{8}}; \nonumber\\
\mathcal{R}_{14}&=&\frac{b_1(b^2_4-b_6b_2)+b_3(b_3b_6-b_4b_5)-b_5(b_3b_4-b_2b_5)}{(ka_{\perp})^{12}};
\label{2}
\end{eqnarray}
where $b_i$ with $i=1,\ldots,6$ are constants given in the following:
\begin{eqnarray}
b_1&=&\zeta\bigg(\frac{1}{2},\epsilon\bigg);~~b_2= 180 \zeta\bigg(-\frac{3}{2},\epsilon\bigg)+30(ka_{\perp})^2\zeta\bigg(-\frac{1}{2},\epsilon\bigg)
+\frac{5}{4}(ka_{\perp})^4\zeta\bigg(\frac{1}{2},\epsilon\bigg)\nonumber\\
b_3&=&2\sqrt{5} \bigg[3\zeta\bigg(-\frac{1}{2},\epsilon\bigg)+\frac{(ka_{\perp})^2}{4}\zeta\bigg(\frac{1}{2},\epsilon\bigg)\bigg] \nonumber\\
b_4&=&6\sqrt{5}\bigg[210\zeta\bigg(-\frac{5}{2},\epsilon\bigg)+\frac{125(ka_{\perp})^2}{2}\zeta\bigg(-\frac{3}{2},\epsilon\bigg)
+\frac{39(ka_{\perp})^4}{8}\zeta\bigg(-\frac{1}{2},\epsilon\bigg)\nonumber\\
&+&\frac{3(ka_{\perp})^6}{24}\zeta\bigg(\frac{1}{2},\epsilon\bigg)\bigg]\nonumber\\
b_5&=&210\zeta\bigg(-\frac{3}{2},\epsilon\bigg)+45(ka_{\perp})^2\zeta\bigg(-\frac{1}{2},\epsilon\bigg)
+\frac{3(ka_{\perp})^4}{8}\zeta\bigg(\frac{1}{2},\epsilon\bigg)\nonumber\\
b_6&=&90\bigg[490\zeta\bigg(-\frac{7}{2},\epsilon\bigg)+210(ka_{\perp})^2\zeta\bigg(-\frac{5}{2},\epsilon\bigg)
+\frac{555(ka_{\perp})^4}{20}\zeta\bigg(-\frac{3}{2},\epsilon\bigg)\nonumber\\
&+&\frac{45(ka_{\perp})^6}{40}\zeta\bigg(-\frac{1}{2},\epsilon\bigg)+\frac{9(ka_{\perp})^8}{640}\zeta\bigg(\frac{1}{2},\epsilon\bigg)\bigg],
\label{3}
\end{eqnarray}
here $\zeta(\cdot,\cdot)$ denotes the Hurwitz zeta function and $\epsilon=3/2-(ka_{\perp}/2)^2$.

Similarly, the scalar product $\boldsymbol{\alpha \cdot \xi}$ results in the following expression:
\begin{eqnarray}
\boldsymbol{\alpha \cdot \xi}&=&\bar{a}_{ss}\xi_{1}+\bar{a}_{dd}\xi_{2}+\bar{a}_{gg}\xi_{3}+\bar{a}_{sd}\xi_{4}+\bar{a}_{dg}\xi_{5}+
\bar{a}_{sd}^2\xi_{6}+\bar{a}_{dg}^2\xi_{7}+\bar{a}_{ss}\bar{a}_{gg}\xi_{8} \nonumber\\
&+&\bar{a}_{ss}\bar{a}_{dg}\xi_{9}+\bar{a}_{ss}\bar{a}_{dd}\xi_{10}+\bar{a}_{gg}\bar{a}_{sd}\xi_{11}+\bar{a}_{dd}\bar{a}_{gg}\xi_{12}+
\bar{a}_{sd}\bar{a}_{dg}\xi_{13} \nonumber\\
&+&(\bar{a}_{ss}\bar{a}_{dg}^2+\bar{a}_{gg}\bar{a}_{sd}^2-\bar{a}_{ss}\bar{a}_{dd}\bar{a}_{gg})\xi_{14},
 \label{4}
\end{eqnarray}
with $\xi_i$, $i=1,\ldots,14$, being the components of the vector $\boldsymbol{\xi}$ and are defined as follows:

\begin{eqnarray}
\xi_{1}&=&2;~~\xi_{2}=\frac{10}{(ka_{\perp})^4};~~\xi_{3}=\frac{81}{2(ka_{\perp})^8};~~\xi_{4}=\frac{4\sqrt{5}}{(ka_{\perp})^2};~~\xi_{5}=\frac{18\sqrt{5}}{(ka_{\perp})^6};
~~\xi_{6}=\frac{-2b_2+4\sqrt{5}b_3-10b_1}{(ka_{\perp})^4}; \nonumber\\
\xi_{7}&=&\frac{18\sqrt{5}b_4-10b_6-\frac{81}{2}b_2}{(ka_{\perp})^{12}};~~\xi_{8}=\frac{\frac{81}{2}b_1-18b_5+2b_6}{(ka_{\perp})^8};
~~\xi_{9}=\frac{18\sqrt{5}b_1-18b_3+4b_4-4\sqrt{5}b_5}{(ka_{\perp})^6}; \nonumber\\
\xi_{10}&=&\frac{10b_1+b_2-4\sqrt{5}b_3}{(ka_{\perp})^4};~~\xi_{11}=\frac{81b_3-18b_4-18\sqrt{5}b_5+4\sqrt{5}b_6}{(ka_{\perp})^{10}};
~~\xi_{12}=\frac{\frac{81}{2}b_2-18\sqrt{5}b_4+10b_6}{(ka_{\perp})^{12}}; \nonumber\\
\xi_{13}&=&\frac{-18b_2+18\sqrt{5}b_3+4\sqrt{5}b_4-20b_5}{(ka_{\perp})^8}; \nonumber\\
\xi_{14}&=&\bigg[\frac{81}{2}(b_3^2-b_1b_2)+18(b_2b_5-b_3b_4)+2(b_4^2-b_2b_6)+4\sqrt{5}(b_3b_6-b_4b_5)+10(b_5^2-b_1b_6) \nonumber\\
&+&18\sqrt{5}(b_1b_4-b_3b_5)\bigg]/(ka_{\perp})^{12},
\label{5}
\end{eqnarray}
with the constants $b_i$, $i=1,\ldots,6$, given in Eq.~(\ref{3}).

Note that in the low-energy regime, $q_0a_{\perp}\ll1$, the dimensionless total energy becomes $\frac{(ka_{\perp})^2}{2}=1+\frac{(q_0a_{\perp})^2}{2}\approx1$.

\section{II. The $\eta$- and $\sigma$-constants of the resonance condition}
The explicit form of the $\eta$- and $\sigma$-constants expressed in terms of the components of $\boldsymbol{\mathcal{R}}$ take on the following appearance:
\begin{eqnarray}
\eta_1&=&-\frac{2\mathcal{R}_2}{21}-\frac{2\mathcal{R}_3}{77}-\frac{\sqrt{5}}{105}(7\mathcal{R}_4+\mathcal{R}_5)\nonumber\\
\eta_2&=&\frac{2\sqrt{5}\mathcal{R}_{11}}{1155}+\frac{4\mathcal{R}_{12}}{1617}+\frac{1}{2205}(7\mathcal{R}_{13}+49\mathcal{R}_6+\mathcal{R}_7)\nonumber\\
\eta_3&=&-\frac{2\mathcal{R}_{14}}{3465}\nonumber\\
\sigma_0&=&\mathcal{R}_1\nonumber\\
\sigma_1&=&-\frac{2\mathcal{R}_{10}}{21}-\frac{2\mathcal{R}_8}{77}-\frac{\sqrt{5}}{105}\mathcal{R}_9\nonumber\\
\sigma_2&=&-\frac{\mathcal{R}_{14}}{495}
\label{6}
\end{eqnarray}

\section{III. The effective 1D Hamiltonian and the transmission coefficient $T$}
The {\it{physical}} $\underline{K}^{1D}$ matrix and the corresponding scattering wave function in the asymptotic regime, $|\mathbf{r}|\to \infty$, are related as follows:
\begin{equation}
 \Psi^{\text{phys}}(z,\rho)=\big[\cos(q_0|z|)-\underline{\tilde{K}}^{1D}_{oo}\sin(q_0|z|)\big]\Phi_{0,0}(\rho)
\label{6}
\end{equation}
where $\Psi^{\text{phys}}$ is the physical wave function \cite{gian12} and $\Phi_{0,0}(\rho)$ is the eigenfunction of the two-dimensional harmonic oscillator for $n=m=0$, i.e. the lowest transversal channel.
Note that $\Psi^{\text{phys}}$ consists only of the real part of the total wave function in the asymptotic regime.
Thus in order to relate the transmission coefficient $T$ to the $\underline{\tilde{K}}^{1D}_{oo}$ matrix we map the relative Hamiltonian $H$ of the two dipoles in the waveguide onto an effective one-dimensional (1D) Hamiltonian where they interact via a 1D delta function, $V(z)=g_{1D}\delta(z)$.
This is permitted since the relative Hamiltonian $H$ asymptotically is fully separable, and all the relevant scattering information is imprinted in the $z$-component of the wave function.
Consequently, by imposing the boundary condition of the $\delta$ function we determine the prefactor $g_{1D}$ such that the real part of the wave function of the effective 1D Hamiltonian is the same as the $z$-dependent part of $\Psi^{\text{phys}}$.
Thus the effective coupling constant $g_{1D}$ becomes $ g_{1D}=-\frac{\hbar^2q_0}{\mu} \underline{\tilde{K}}_{oo}^{1D}$ and the corresponding transmission coefficient $T$ of the effective 1D Hamiltonian reads
\begin{equation}
 T=\frac{1}{1+\frac{\mu^2g_{1D}^2}{\hbar^4q_0^2}}=\frac{1}{1+(\underline{\tilde{K}}^{1D}_{oo})^2}.
\label{7}
\end{equation}